\documentclass[10pt]{article} 
\usepackage[english]{babel}
\usepackage[utf8]{inputenc}
\usepackage{geometry}
\usepackage{mathtools}
\usepackage{amsmath}
\usepackage{amsfonts}
\usepackage{hyperref}
\usepackage[ruled,vlined]{algorithm2e}
\usepackage{caption}
\usepackage{subcaption}
\usepackage{bm}
\usepackage{bbm}
\usepackage{amssymb}

\usepackage{graphicx}
\usepackage{setspace}
\usepackage{adjustbox}
\usepackage[backend=bibtex, style=authoryear, maxcitenames=2, maxbibnames=99, dashed=false]{biblatex}
\usepackage [english]{babel}

\begin{document}

\title{Particle Hamiltonian Monte Carlo}

\author{Alaa Amri$^{a}$ \footnote{Corresponding author: Alaa, Amri; \tt{alaa.amri@ed.ac.uk}} , V\'{i}ctor Elvira$^{a}$ and Amy L. Wilson$^{a}$.\\
        \small $^{a}$School of Mathematics, The University of Edinburgh (United Kingdom)\\
}

\date{} 

\maketitle

\begin{abstract} 
\noindent In Bayesian inference, Hamiltonian Monte Carlo (HMC) is a popular Markov Chain Monte Carlo (MCMC) algorithm known for its efficiency in sampling from complex probability distributions. However, its application to models with latent variables, such as state-space models, poses significant challenges. These challenges arise from the need to compute gradients of the log-posterior of the latent variables, and the likelihood may be intractable due to the complexity of the underlying model.
In this paper, we propose Particle Hamiltonian Monte Carlo (PHMC), an algorithm specifically designed for state-space models. PHMC leverages Sequential Monte Carlo (SMC) methods to estimate the marginal likelihood, infer latent variables (as in particle Metropolis-Hastings), and compute gradients of the log-posterior of model parameters. Importantly, PHMC avoids the need to calculate gradients of the log-posterior for latent variables, which addresses a major limitation of traditional HMC approaches.
We assess the performance of Particle HMC on both simulated datasets and a real-world dataset involving crowdsourced cycling activities data. The results demonstrate that Particle HMC outperforms particle marginal Metropolis-Hastings with a Gaussian random walk, particularly in scenarios involving a large number of parameters.

 \end{abstract}

\begin{flushleft}
\end{flushleft}

\section{Introduction}
The Metropolis-Hastings algorithm relies on selecting an appropriate proposal distribution to efficiently sample the target distribution. The choice of proposal impacts the acceptance rate and how effectively the Markov chain explores the parameter space. A poorly chosen proposal can result in slow convergence and high rejection rates. While random walk proposals are common and yield high acceptance rates for small steps, they can be inefficient in high-dimensional settings, causing low acceptance rates, slow mixing, and highly correlated samples. Therefore, careful design of the proposal distribution is key to ensuring accuracy and efficiency. \newline
    
Significant advancements were achieved when the Metropolis Adjusted Langevin Algorithm (MALA) was introduced, which utilizes a proposal process derived from a discretized Langevin diffusion that incorporates gradient information of the target density in its drift term \parencite{roberts1996exponential}. Similarly, the Hamiltonian Monte Carlo (HMC) method, originally proposed in the field of statistical physics by \textcite{duane1987hybrid} to efficiently simulate states of physical systems, was later adapted for statistical inference problems \parencite{neal1993probabilistic,liu2001monte}. HMC is a Markov chain Monte Carlo (MCMC) algorithm for sampling from probability distributions. It can propose moves that are distant from the chain's current state but still have a high probability of being accepted. The method is based on simulating a Hamiltonian system and involves two key parameters: the duration of the Hamiltonian flow and the integrator's time step size. However, calculating gradients can be difficult, and in some cases, they are replaced by estimates, a technique referred to as "stochastic gradients". Two well-known examples of this approach are stochastic gradient Langevin dynamics (SGLD) \parencite{welling2011bayesian} and stochastic gradient Hamiltonian Monte Carlo (SGHMC) \parencite{chen2014stochastic}.\newline

In Metropolis-Hastings algorithms, replacing the intractable likelihood with an estimate gives rise to pseudo-marginal MCMC methods \parencite[]{andrieu2009pseudo}, which allow for Bayesian inference even when the likelihood is difficult or impossible to compute directly. In cases where sequential Monte Carlo (SMC) methods are used to provide this estimate, the resulting algorithm is referred to as particle MCMC \parencite{pmcmc}. It is worth noting that there are particle MCMC algorithms based on MALA \parencite{dahlin2013particle,nemeth2016particle,corenflos2024particle}.
\textcite{alenlov2021pseudo} proposed pseudo-marginal Hamiltonian Monte Carlo (HMC) where the intractable likelihood is replaced by an unbiased estimate, though their method does not account for gradient estimation and the unbiased estimate is not obtained by sequential Monte Carlo methods. \textcite{Osmundsen2018} built upon their work, to apply it to dynamic systems.

\subsection{Objective and plan}
\begin{flushleft}
 In this paper, our goal is to develop a particle marginal Metropolis-Hastings algorithm where parameter proposals are generated using Hamiltonian dynamics, as in Hamiltonian Monte Carlo (HMC). An SMC method is incorporated to estimate gradients, compute the marginal likelihood, and infer latent variables. This combined approach, referred to as particle Hamiltonian Monte Carlo, is designed for state-space models.

The remainder of the paper is structured as follows: Section 2 provides a review of the Hamiltonian Monte Carlo method, while Section 3 revisits two gradient estimation techniques based on SMC methods. In Section 4, we propose Particle Hamiltonian Monte Carlo. In Section 5, we present a numerical demonstration of the proposed method using datasets simulated from two different state-space models and real dataset of crowdsourced cycling data. We conclude the paper with Section 6.

\section{Bayesian Inference for State-Space Models}
\begin{flushleft}
Let us consider a state-space model which can be expressed as follows   
\end{flushleft}
\begin{equation} 
\label{ssmodel}
 \begin{split}
     Y_{t}|H_{t} = h_{t}  &\sim p_\theta(y_t \mid h_t), \\
     H_{t} \mid H_{t-1} = h_{t-1}  &\sim  p_\theta(h_t \mid h_{t-1}),\\
      H_{1} &\sim p_\theta(h_1), 
 \end{split}
\end{equation}

\begin{flushleft}
    where we consider two stochastic processes: \(\{ H_t \}_{t=1}^T\), taking values in \(\mathcal{X}\), representing hidden variables, and \(\{ Y_t \}_{t=1}^T\), taking values in \(\mathcal{Y}\), representing observations. The model parameters are denoted by \(\theta \in \Theta\). The initial distribution, \(p_\theta(h_1)\), defines the prior for the hidden state \(H_1\) at \(t=1\). The transition distribution, \(p_\theta(h_t \mid h_{t-1})\), describes the evolution of latent variables between time steps \(t-1\) and \(t\) in the absence of observations, while \(p_\theta(y_t \mid h_t)\) specifies the probability of observing \(Y_t\) given \(H_t = h_t\). At each time step \(t\), \(h_t\) and \(y_t\) are realizations of the random variables \(H_t\) and \(Y_t\), respectively. This work focuses on inference for both the model parameters \(\theta\) and the latent variables \(H_{1:T}\).
\end{flushleft}

\subsection{Sequential importance sampling with resampling}
\begin{flushleft}
    Sequential Monte Carlo methods provide a framework for inferring latent variables \( H_{1:T} \). A well-known technique within this framework is sequential importance sampling with resampling. At each time step \( t \), this method samples particles and calculates their corresponding weights \( w_t \), which are normalized and stored to estimate target quantities or approximate posterior distributions, such as filtering distributions. Normalized weights are represented by \( W_t \). Resampling can be performed adaptively when the variability in weights becomes too large. This strategy, known as adaptive resampling, is initiated when the Effective Sample Size (ESS) falls below a specified threshold \( ESS_{min} \). Algorithm \ref{alg1.2} outlines the pseudo-code for a generic particle filter, where operations with the superscript \( i \) are executed individually for each of the \( N \) particles.
\end{flushleft}

\begin{algorithm}[!ht]
\DontPrintSemicolon
\tcp{Operations involving the superscript $i$ are performed for $i = 1,\dots,N$.}

Sample $H_1^{i} \sim q_{\theta}(.)$\;

Compute weights \[
w_{1}^i = \frac{p_\theta\left(H_{1}^{i}\right) p_\theta\left(y_1 \mid H_{1}^{i}\right)}{q_{\theta}\left(H_{1}^{i}\right)}.
\]

Normalize weights \[
W_{1}^i = \frac{w_{1}^i}{\sum_{j=1}^{N} w_{1}^j}.
\]

\For{$t = 2,\dots,T$}{
    Calculate $\text{ESS} := \frac{1}{\sum_{i=1}^{N} \left(W_{t-1}^{i}\right)^{2}}$\;
    
    \If{$\text{ESS} \leq \text{ESS}_{\text{min}}$}{
        Draw the index $a_{t}^{i}$ of the ancestor of the particle $i$, by resampling the normalized weights $W_{t-1}^{1:N}$\;
        Set $\hat{w}^{i}_{t-1} = 1$\;
    }
    \Else{
        $a_{t}^{i} = i$\;
        $\hat{w}^{i}_{t-1} = w_{t-1}^{i}$\;
    }

    Sample $H_t^{i} \sim q_{\theta}(. \mid H_{t-1}^{a_{t}^{i}})$\;

    Compute weights \[
    w_{t}^i = \frac{p_\theta\left(H_{t}^{i} \mid H_{t-1}^{a_{t}^{i}}\right) p_\theta\left(y_t \mid H_{t-1}^{a_{t}^{i}}\right)}{q_{\theta}\left(H_{t}^{i} \mid H_{t-1}^{a_{t}^{i}}\right)} \hat{w}^{i}_{t-1}.
    \]

    Normalize weights \[
    W_{t}^i = \frac{w_{t}^i}{\sum_{j=1}^{N} w_{t}^j}.
    \]
}
\caption{Sequential importance sampling with resampling}
\label{alg1.2}
\end{algorithm}

\begin{flushleft}
    In our paper, we employed systematic resampling, which empirically outperforms other resampling methods like stratified resampling and particularly multinomial resampling. It produces estimates with reduced variability, making it more effective. \parencite[Chapter 9]{Chopin2020}.
    Furthermore, let us define $\ell_t$ which estimates $p_{\theta}(y_{t}\mid y_{1:t-1})$ for $t > 1$ and $\ell_1$ that estimates  $p_{\theta}(y_{1})$, they are both defined as follows 
    \begin{equation}
\begin{split}
   \ell_1 &= \frac{1}{N} \sum_{i=1}^N w_1^i,\\
    \ell_t &=\left\{\begin{array}{l}\frac{1}{N} \sum_{i=1}^N w_t^i \text { if resampling was performed at time } t>1, \\ \frac{\sum_{i=1}^N w_t^i}{\sum_{i=1}^N w_{t-1}^i} \text { otherwise }\end{array}\right.
\end{split}
\end{equation}
\end{flushleft}

\begin{flushleft}
    Subsequently, it is easy to establish that $p_{\theta}(y_{1:T})$ can be estimated by $\hat{p}_{\theta}(y_{1:T})$ in the following way
\begin{equation}
   p_{\theta}(y_{1:T}) = p_{\theta}(y_{1}) \prod_{t=2}^T p_{\theta}(y_{t}\mid y_{1:t-1}) \approx  \hat{p}_{\theta}(y_{1:T}) = \ell_1 \prod_{t=2}^T \ell_t.
\end{equation}
It should be noted that the estimate of marginal likelihood $p_{\theta}\left(y_{1:T} \right)$ is unbiased  \parencite{delmoral}. \textcite{delmoral2} showed that the estimate $\hat{p}_{\theta^{}}\left(y_{1: T}\right)$ has a non-asymptotic variance that grows linearly with the time horizon $T$. 
\end{flushleft}

\subsection{Particle Marginal Metropolis-Hastings}
\begin{flushleft}
    The particle marginal Metropolis-Hastings algorithm with $K$ iterations, broadly consists of using a sequential Monte Carlo method at each iteration to sample the latent variables and using the unbiased marginal likelihood estimate in the acceptance probability. A generic choice of the proposal used to sample parameters $\theta$ is a Gaussian random walk. The method is presented in Algorithm \ref{pmmh_alg}.
\end{flushleft}

\begin{algorithm}[h]
\DontPrintSemicolon

\textbf{Input:} $K, N, \theta^{(0)}$\\

Run an SMC algorithm targeting $p_{\theta^{(0)}}\left(h_{1: T} \mid y_{1: T}\right)$, store a sample $H_{1: T}^{(0)}$ and let $\hat{p}_{\theta^{0}}\left(y_{1: T}\right)$ denote the marginal likelihood estimate. \\
  \For{k = 1,..,K}{

Sample $\theta^{'} \sim$  $q \left( . \mid \theta^{(k-1)}\right)$,\\
Run an SMC algorithm targeting $p_{\theta^{'}}\left(h_{1: T} \mid y_{1: T}\right)$, store a sample $H_{1: T}^{'}$ and let $\hat{p}_{\theta^{'}}\left(y_{1: T}\right)$ denote the marginal likelihood estimate.\\
With probability,
  \\
$$
1 \wedge \frac{\hat{p}_{\theta^{'}} \left(y_{1: T}\right) p\left( \theta^{'}\right) q\left(\theta^{(k-1)} \mid \theta^{'}\right)}{\hat{p}_{\theta^{(k-1)}}\left(y_{1: T}\right) p\left(\theta^{(k-1)}\right) q\left(\theta^{'} \mid \theta^{(k-1)}\right)}.
$$ 
  \\
Set $\theta^{(k)}=\theta^{'}$, $H_{1: T}^{k}=H_{1: T}^{'}$ and $\hat{p}_{\theta^{(k)}}\left(y_{1: T}\right)=\hat{p}_{\theta^{'}}\left(y_{1: T}\right)$; otherwise set $\theta^{(k)}=\theta^{(k-1)}$, $H_{1: T}^{k}=H_{1: T}^{k-1}$ and $\hat{p}_{\theta^{(k)}}\left(y_{1: T}\right)=\hat{p}_{\theta^{(k-1)}}\left(y_{1: T}\right)$.
}

\caption{Particle Marginal Metropolis–Hastings}
\label{pmmh_alg}
\end{algorithm}
\end{flushleft}

\section{Hamiltonian Monte Carlo}
\begin{flushleft}
    The Hamiltonian Monte Carlo (HMC) is a general purpose Markov chain Monte Carlo (MCMC) algorithm for sampling from a probability distribution of the form $\pi (\theta) \propto  \exp (-U(\theta))$
    where the potential energy function $U: \mathbb{R}^{d_{\theta}} \rightarrow \mathbb{R}$ is assumed to be twice continuously differentiable.
    We introduce auxiliary momentum variables $r$ (where $r \in \mathbb{R}^{d_{\theta}}$) to define the Hamiltonian function or the total energy function, $Ha: \mathbb{R}^{2d_{\theta}} \rightarrow \mathbb{R}$, that is the sum of the potential function and the kinetic function $\frac{1}{2} r^\top  \mathbf{M}^{-1} r$:\\
    \begin{equation*}
Ha(\theta, r) =   U(\theta)   + \frac{1}{2} r^\top  \mathbf{M}^{-1} r.
\end{equation*}
    It is a common assumption that the  momentum variables $r \sim \mathcal{N} (\mathbf{0}, \mathbf{M} )$, where $\mathbf{M}$ represents a symmetric, positive-definite matrix known as the mass matrix, often simplified by setting it to the $d_{\theta} \times d_{\theta}$ identity matrix $\mathbf{I}$ \parencite[Chapter 5]{neal1993probabilistic}. An appropriately selected mass matrix enhances algorithm effectiveness, enabling the sampler to navigate the parameter space more efficiently. This, in turn, substantially increases the acceptance rate of proposed states. However, there is no established guideline for selecting the best mass matrix given a certain model. \\ The evolution of the trajectory of $(\theta, r)$ under the Hamiltonian dynamics is given by the following ordinary differential equations
\begin{equation} \begin{aligned}
    \begin{array}{l}
\frac{d \theta}{d \tau}=\frac{\partial Ha}{\partial r}=\mathbf{M}^{-1} r, \\
\frac{d r}{d \tau}=-\frac{\partial Ha}{\partial \theta}= -\nabla_\theta U(\theta).
\end{array}
\label{eq0}
\end{aligned} \end{equation}
    Note that the derivatives $\frac{d \theta}{d \tau}$ and $\frac{d r}{d \tau}$ are with respect to the continuous time $\tau$ and $\nabla_\theta$ denotes the gradient with respect to $\theta$. As the momentum is equal to mass times velocity, $\mathbf{M}^{-1} r$ is represented as velocity but in our case we considered the case where the mass matrix $\mathbf{M}$ is set to the identity matrix so the derivative of $\theta$ with respect to $\tau$ is simply the momentum variables $r$.
    \end{flushleft}
\begin{flushleft}
    The flow induced by (\ref{eq0}) has the following properties: a) \textit{Conservation of the Hamiltonian}: this indicates that the total energy of the system, as defined by the Hamiltonian function $Ha$, remains unchanged over time, b) \textit{Volume preservation}: Applying the mapping to points within a designated region $R$ in the ($\theta$, $r$) phase space, which has a volume $V$, results in the transformed image of $R$ under the mapping having the same volume, $V$ and c) \textit{time reversibility}:
    The significance of the reversibility of Hamiltonian dynamics lies in its ability to demonstrate that Markov chain Monte Carlo (MCMC) updates employing these Hamiltonian dynamics maintain the desired distribution invariant.

    In practical terms, an exact solution formula of ($\ref{eq0}$) is not available, necessitating the use of a numerical solution instead to discretize the equations using small time step $\epsilon$. One common choice is the St\"{o}rmer-Verlet (also known as leapfrog integrator) \parencite{hairer2003geometric}. On another note, it is also possible to use other ordinary differential equations (ODE) solvers \parencite{hairer2012numerical}. The St\"{o}rmer-Verlet iterates the following updates:
\begin{equation*}
\begin{split}
     r_{\tau + \frac{\epsilon}{2}} &\leftarrow r_{\tau} + \frac{\epsilon}{2}\nabla_{\theta} U\left(\theta \right)|_{\theta=\theta_{\tau}}\\
     \theta_{\tau + \epsilon}  &\leftarrow \theta_{\tau} + \epsilon r_{\tau + \frac{\epsilon}{2}}\\
      r_{\tau + \epsilon} &\leftarrow r_{\tau + \frac{\epsilon}{2}} + \frac{\epsilon}{2} \nabla_{\theta} U\left(\theta \right)|_{\theta=\theta_{\tau + \epsilon}}. \\
\end{split}
\end{equation*}
    We denote by $\theta_{\tau}$ and $r_{\tau}$ the values of the position and momentum variables, respectively, at time $\tau$. The leapfrog integrator functions as a symplectic integrator, maintaining the symplectic structure inherent in Hamiltonian dynamics. This property guarantees that the transformation carried out on the combined space of positions and momenta during leapfrog updates preserves the volume in phase space. In simpler terms, the volume of a given region remains constant when each point in that region is mapped to a new point using the leapfrog integrator. The property of energy conservation of the Hamiltonian dynamics implies that proposals are always accepted. However,
    an additional Metropolis–Hastings accept-reject step is incorporated to correct the bias arising from time discretization error. 
    As a result, the acceptance rate is no longer equal to 1 but it is notable that the acceptance rates remain generally high, even for proposals that may deviate significantly from their previous state. 
    
    In this case, consider a straightforward approach where we integrate the initial state $ (\theta_{0},r_{0})$ over a fixed duration, which is equivalent to $L$ symplectic integrator steps. We then propose the final state of the numerical trajectory $ (\theta_{L},r_{L})$.
    After that, we flip the sign of the momentum which also leaves the total energy unchanged. Let $U(\theta) = - \text{log } p(y,\theta)$, the acceptance probability of the Metropolis-Hastings step would be given as follows
\begin{equation}
\begin{split}
    &min ( 1, \frac{exp(-Ha(\theta_{L},-r_{L}))}{exp(-Ha(\theta_{0},r_{0}))}  ) = min( 1, exp(-Ha(\theta_{L},-r_{L})+Ha(\theta_{0},r_{0})) ) \\
    = &min\left(1, \exp\left(  log\left(p_{} \left(y \mid \theta_{L} \right) p\left(\theta_{L}\right) \right) - log\left(p_{} \left(y \mid \theta_{0} \right) p\left(\theta_{0}\right) \right)  + \frac{1}{2} r_{0}^\top  r_{0}  - \frac{1}{2} r_{L}^\top  r_{L} \right)\right).
\end{split}
\end{equation}
\end{flushleft}

\begin{algorithm}[h]
\DontPrintSemicolon
\textbf{Input:} $K$, $L$, $\theta^{(0)}, \epsilon, \mathcal{L}$\\

  \For{k = 1,..,K}{
Sample $r_{0} \sim \mathcal{N} \left(0, I \right)$.\\
Set $\theta_{0} \leftarrow \theta^{(k-1)}$.\\

\For{i = 1,..,L}{
Set $r_{i} \leftarrow r_{i-1} + \frac{\epsilon}{2}$ $\nabla_{\theta} U\left(\theta \right)|_{\theta=\theta_{i-1}}$.\\

Set $\theta_{i}  \leftarrow \theta_{i-1} + \epsilon r_{i}$.\\

Set $r_{i} \leftarrow r_{i} + \frac{\epsilon}{2}$ $\nabla_{\theta} U\left(\theta \right)|_{\theta=\theta_{i}}$. \\
}

With probability
  \\
$
min\{ 1, exp(-Ha(\theta_{L},-r_{L})+Ha(\theta_{0},r_{0})) \}
$, 
  \\
Set $\left(\theta^{(k)},r^{(k)}  \right) = \left(\theta_{L},-r_{L} \right)$.

  }

\caption{Hamiltonian Monte Carlo}
\end{algorithm}

\section{Estimation of the gradients}
\begin{flushleft}
In this section, we review a way to estimate the gradient of the log-posterior of the model parameters $\nabla_{\theta} \text{log p}(\theta|y_{1:T})$, which has a linear computational cost.
In addition to describing this method, we will critically examine the potential shortcomings of this estimation technique. These may include challenges related to the variance of the gradient estimates. To address these issues, we will describe an alternative approach that offers improvements, which was proposed in \textcite{poy}. 
\end{flushleft}
\subsection{\( O(N) \) Particle-Based Gradient Approximation}
 \begin{flushleft}
    HMC and MALA make use of the gradient of the log-posterior of the model's parameters $\nabla_{\theta} \text{log p}(\theta|y_{1:T})$ which can be expressed as the sum of the gradient of the logarithm of the parameter's prior density $\nabla_{\theta} \text{log p}(\theta)$, where it is assumed that it can be calculated explicitly, and the gradient of the log-likelihood, known as the score vector $\nabla_{\theta} \text{log p}_{\theta}(y_{1:T})$. 
\begin{equation} \begin{aligned}
\label{eq1}
		\nabla_{\theta} \text{log p}(\theta|y_{1:T}) = \nabla_{\theta} \text{log p}_{\theta}(y_{1:T}) + \nabla_{\theta} \text{log p}(\theta).
\end{aligned} \end{equation}
    Fisher's identity \parencite{int1} allows us to express the score vector as an expectation and it assumes that the density $\text{p}_{\theta}(y_t|h_{t})$ and the state transition density $\text{p}_{\theta}(h_t|h_{t-1})$ exist in addition to admitting tractable first and second derivatives. It also assumes that all functions are regular enough so that an exchange between derivation and integration can be performed. (see proof in the Appendix \ref{appendix})
\begin{equation} \begin{aligned}
\label{eq2}
\nabla_{\theta} \text{log p}_{\theta}(y_{1:T}) &= \int 
  \text{p}_{\theta}(h_{1:T}|y_{1:T}) \nabla_{\theta} \text{log p}_{\theta}(y_{1:T}, h_{1:T}) dh_{1:T} \\
  &= E[ \nabla_{\theta} \text{log p}_{\theta}(y_{1:T}, h_{1:T})|Y_{1:T}=y_{1:T}].
\end{aligned} \end{equation} 
    Hence, it is enough to have a particle approximation of $\text{p}_{\theta}(h_{1:T}|y_{1:T})$ in order to obtain an approximation of $\nabla_{\theta} \text{log p}_{\theta}(y_{1:T})$. The gradient of the logarithm of the joint density $\text{log p}_{\theta}(y_{1:T}, h_{1:T})$ can be expressed in an additive way as 
\begin{equation} \begin{aligned}
\label{eq3}
\begin{split}
 \nabla_{\theta} \text{log p}_{\theta}(y_{1:T}, h_{1:T}) &= 
  \nabla_{\theta} \text{log p}_{\theta}(h_1) + \sum_{t=2}^{T} \nabla_{\theta} \text{log p}_{\theta}(h_t|h_{t-1})+ \sum_{t=1}^{T} \nabla_{\theta} \text{log p}_{\theta}(y_t|h_{t}) \\
  &= \nabla_{\theta} \text{log p}_{\theta}(y_{1:T-1}, h_{1:T-1}) + 
  \nabla_{\theta} \text{log p}_{\theta}(h_T|h_{T-1}) + 
  \nabla_{\theta} \text{log p}_{\theta}(y_T|h_{T}).
\end{split}
\end{aligned} \end{equation} 
\end{flushleft}
\begin{flushleft}
    Let $g_{1}^{(i)}\left(\theta\right)$, $\Tilde{g}_{t}^{(i,j)}\left(\theta\right)$ and $g_{t}^{(i,j)}\left(\theta\right)$ be the estimates of $\nabla_{\theta} \log p_\theta\left(y_1 \mid h_1^{}\right)$, $\nabla_{\theta} \text{log p}_{\theta}(y_{t}, h_{t} \mid h_{t-1})$, and $\nabla_{\theta} \text{log p}_{\theta}(y_{1:t}, h_{1:t})$ respectively. Here, \( i \) and \( j \) represent the indices of two particles out of the \( N \) particles sampled in the particle filter. These estimates are expressed as
\begin{equation} \begin{aligned}
\label{eq4}
g_{1}^{(i)}\left(\theta\right) &= \nabla_{\theta} \log p_\theta\left(y_1 \mid H_1^{i}\right)+\nabla_{\theta} \log p_\theta\left(H_1^{i}\right), \\
g_{t}^{(i,j)}\left(\theta\right) &=g_{t-1}^{(j,j)}\left(\theta\right)+ \underbrace{\nabla_{\theta} \log p_\theta\left(y_t \mid H_t^{i}\right)+\nabla_{\theta} \log p_\theta\left(H_t^{i} \mid H_{t-1}^{j}\right)}_\text{$\Tilde{g}_{t}^{(i,j)}\left(\theta\right)$}\quad\text { for }t = 2,..,T , \\
g_{t}^{(i,i)}\left(\theta\right)  &= g_{1}^{(i)}\left(\theta\right) + \sum_{k=2}^{t} \Tilde{g}_{k}^{(i,i)}\left(\theta\right).
\end{aligned} \end{equation}
    Consequently, we can construct the estimates of both $\nabla_{\theta} \text{log p}_{\theta}(y_{1:T})$ and $\nabla_{\theta} \text{log p}(\theta|y_{1:T})$ which are denoted respectively by $S_{T}\left(\theta\right)$ and $G_{T}\left(\theta\right)$:
\begin{equation}
S_T\left(\theta\right)  =\sum_{i=1}^N W_T^{i}(\theta) g_{T}^{(i,i)}\left(\theta\right), \quad \text { and } \quad G_T\left(\theta\right)  =S_T\left(\theta\right)  + \nabla_{\theta} \text{log p}(\theta).
\end{equation}
\end{flushleft}

\subsection{Online \( O(N^2) \) Particle-Based Gradient Approximation Method}
\begin{flushleft}
The results are not introduced here but \textcite{poy} demonstrates that, even with appealing mixing assumptions, the variance of this estimate grows at least quadratically with time $T$ as the particle approximation of $\text{p}_{\theta}(h_{1:T}|y_{1:T})$ becomes progressively degraded due to successive resampling steps. Specifically, the number of distinct particles approximating  for any fixed $m<T$ decreases as $T-m$ increases.
The resampling step inherently causes the particle filter approximation to consist of a single particle at some time $m$. With a fixed number of particles $N$, it becomes impossible to accurately approximate $\text{p}_{\theta}(h_{1:T}|y_{1:T})$ when $T$ is large which renders the method ineffective for long sequences.
This well-known problem in the literature is referred to as the path degeneracy problem. \parencite{poy}  suggested a method does not suffer from path degeneracy at the cost of $O(N^2)$ as it computes for every particle (given that we have $N$ particles) a sum of $N$ terms. On the other hand, it is possible to obtain estimates with variance that grows linearly with time $T$ (not at least quadratically like the previous approximation method that has $O(N)$ cost). The method is described in detail herein.

\end{flushleft}
\begin{flushleft}
    Consider the distribution of $h_{t}$ given both $h_{t+1}$ and $Y_{1:t}$, it can be expressed as follows using the Bayes' rule:
\begin{equation} \begin{aligned}
    p(h_{t} | h_{t+1} , y_{1:t}) = \frac{p(h_{t+1} | h_{t}) p(h_{t} | y_{1:t})}{\underbrace{\int p(h_{t+1} | h_{t}) p(h_{t} | y_{1:t}) dh_{t}}_\text{$p(h_{t+1} | y_{1:t})$}}.
    \label{eq5}
\end{aligned} \end{equation}
    Assuming that a particle filter was run and the weights were calculated, the filtering distribution $p(h_{t} | y_{1:t})$ can be approximated as:
\begin{equation} \begin{aligned}
     p(h_{t} | y_{1:t}) \approx \sum_{i=1}^{N} W_t^{i}(\theta) \delta_{H_{t}^{i}} (h_{t}) .
     \label{eq6}
\end{aligned} \end{equation}
    Where $\delta(.)$ is the Dirac delta function. Hence, we can get an approximation for $p(h_{t} | h_{t+1} , y_{1:t})$ by plugging $\ref{eq6}$ into $\ref{eq5}$
\begin{equation} \begin{aligned}
    \begin{split}
        p(h_{t} | h_{t+1} , y_{1:t}) &\approx \frac{p(h_{t+1} | h_{t}) \sum_{i=1}^{N} W_t^{i} (\theta)\delta_{H_{t}^{i}} (h_{t})}{\int p(h_{t+1} | h_{t}) \sum_{i=1}^{N} W_t^{i}(\theta) \delta_{H_{t}^{i}} (h_{t}) dh_{t}} \\
        &= \frac{\sum_{i=1}^{N} W_t^{i}(\theta) p(h_{t+1} | H_{t}^{i}) \delta_{H_{t}^{i}} (h_{t})}{\sum_{i=1}^{N} W_t^{i}(\theta) p(h_{t+1} | H_{t}^{i})}.\\
    \end{split}
\end{aligned} \end{equation}
    Consequently, we can obtain an approximation of any expectation with respect to $p(h_{t} | h_{t+1} , y_{1:t})$ in the following way
\begin{equation} \begin{aligned}
    E[ \phi(h_{t})|H_{t+1} = h_{t+1}, Y_{1:t}=y_{1:t}] \approx \frac{\sum_{i=1}^{N} W_t^{i}(\theta) p(h_{t+1} | H_{t}^{i}) \phi ( H_{t}^{i})}{\sum_{i=1}^{N} W_t^{i}(\theta) p(h_{t+1} | H_{t}^{i})}.
\end{aligned} \end{equation}
    By the Markov property, and for $t$ $\leq$ $T$ and any function $\phi$ we have 
    \begin{equation} \begin{aligned}
    E[ \phi(h_{t-1})|H_{t} = h_{t}, Y_{1:t}=y_{1:t}] =
E[ \phi(h_{t-1})|H_{t} = h_{t}, Y_{1:T}=y_{1:T}].
\end{aligned} \end{equation}
\end{flushleft}

\begin{flushleft}
    Hence, $E[ \phi(h_{t-1})|Y_{1:t}=y_{1:t}]
        $ can be approximated as
\begin{equation} 
    \begin{split}
       E[ \phi(h_{t-1})|Y_{1:t}=y_{1:t}]
        &= E[ E[ \phi(h_{t-1})|H_{t} = h_{t}, Y_{1:T}=y_{1:T}] | Y_{1:t}=y_{1:t}]  \\
        &\approx \sum_{j=1}^{N} W_t^{j}(\theta) \frac{\sum_{i=1}^{N} W_{t-1}^{i}(\theta) p(H_{t}^{j} | H_{t-1}^{i}) \phi ( H_{t-1}^{i})}{\sum_{i=1}^{N} W_{t-1}^{i}(\theta) p(H_{t}^{j} | H_{t-1}^{i})}.\\
    \end{split}
\end{equation}
    Let us consider the case where $\phi ( H_{t-1}^{i}) = g_{t}^{(j,i)}\left(\theta\right)$ then we define $S^{'}_t\left(\theta\right)$ as
\begin{equation}
    S^{'}_t\left(\theta\right) = \sum_{j=1}^{N} W_t^{j}(\theta) \frac{\sum_{i=1}^{N} W_{t-1}^{i}(\theta) p(H_{t}^{j} | H_{t-1}^{i}) g_{t}^{(j,i)}\left(\theta\right)}{\sum_{i=1}^{N} W_{t-1}^{i}(\theta) p(H_{t}^{j} | H_{t-1}^{i})}.
\end{equation}
    Finally, it can be seen that $S^{'}_T\left(\theta\right)$ is also an estimate of $\nabla_{\theta} \text{log p}_{\theta}(y_{1:T})$. We then define $G^{'}_T\left(\theta\right)$, the estimate of $\nabla_{\theta} \text{log p}(\theta|y_{1:T})$, as 
\begin{equation}
    G^{'}_T\left(\theta\right)  =S^{'}_T\left(\theta\right)  + \nabla_{\theta} \text{log p}(\theta).
\end{equation}
\end{flushleft}
 
\section{Proposed algorithm: Particle Hamiltonian Monte Carlo}
\label{phmcsubsection}

\begin{flushleft}
We now propose a modified version of the Hamiltonian Monte Carlo (HMC) method specifically designed for state-space models, with the details provided in Algorithm \ref{phmc_algorithm}. This approach is aimed at efficiently inferring both the latent variables and the unknown parameters of the model.  In traditional HMC, the sampling process involves augmenting the parameter space with artificial "momentum" variables, which are drawn from a standard normal distribution, and then using Hamiltonian dynamics to propose new parameters. Our method retains the core idea of drawing momenta from a standard normal distribution in each iteration of the Markov chain, but adapts the subsequent steps to account for the state-space model structure.

After initializing the momenta, a sequential Monte Carlo algorithm is run using the parameter values sampled from the previous iteration of the algorithm, denoted by \(\theta^{(0)}\). The SMC algorithm provides two essential estimates: (i) the gradient of the log-likelihood with respect to the parameters, denoted as \(-G_T'(\theta^{(0)})\), and (ii) the marginal likelihood estimate \(\hat{p}_{\theta^{(0)}}\left(y_{1:T}\right)\), where \(y_{1:T}\) are the observed data points from time 1 to \(T\). These estimates are essential for guiding the HMC updates.

Following this, the leapfrog integration method is applied, which is a standard procedure in HMC for updating the parameters and momenta over multiple steps. In each leapfrog step, the gradient of the potential energy function (which is typically used to update the momenta) is substituted by the estimate obtained from the SMC algorithm \(-G_T'(\theta^{(i-1)})\). This means that instead of relying on exact gradients of the likelihood (which can be intractable for state-space models), we use the gradient approximation obtained via sequential Monte Carlo (SMC) methods. This cycle of updating parameters and using a SMC algorithm to estimate gradients is repeated until the final step of the leapfrog integration, denoted by the \(L\)-th step. At this final momentum update, the marginal likelihood estimate \(\hat{p}_{\theta^{(L)}}\left(y_{1:T}\right)\) corresponding to the final parameter values \(\theta^{(L)}\) is stored. This marginal likelihood estimate is crucial for computing the acceptance probability in the Metropolis-Hastings step of the HMC algorithm, which is calculated as follows: 
\begin{equation} \begin{aligned}
    &\min\left(1, \exp\left(\hat{Ha}(\theta^{(0)},r_{0}) - \hat{Ha}(\theta^{(L)},r_{L})\right)\right) \\
    &= \min\left(1, \exp\left(  log\left(\hat{p}_{\theta^{(L)}}\left(y_{1: T}\right) p\left(\theta^{(L)}\right) \right) - log\left(
 \hat{p}_{\theta^{(0)}}\left(y_{1: T}\right) p\left(\theta^{(0)}\right) \right)   + \frac{1}{2} r_{0}^\top  r_{0}  - \frac{1}{2} r_{L}^\top  r_{L} \right)\right).
\end{aligned} \end{equation}
    The proposed parameters and the latent variables at the end of the leapfrog integration are accepted with the probability above. By using this modified procedure, the algorithm adapts HMC to the structure of state-space models, leveraging SMC to approximate the necessary gradients and likelihood. 
\end{flushleft}
\begin{algorithm}[h]
\DontPrintSemicolon
\textbf{Input:} $K, L, N, \theta^{(0)}, \epsilon$\\

  \For{k = 1,..,K}{
Sample $r_{0} \sim \mathcal{N} \left(0, I \right)$.\\
Set $\theta^{(0)} \leftarrow \theta^{(k-1)}$.\\

Run an SMC algorithm, using $N$ particles, targeting $p_{\theta^{(0)}}\left(h_{1: T} \mid y_{1: T}\right)$ then store $H_{1: T}^{(0)}$ and also store the marginal likelihood estimate $\hat{p}_{\theta^{(0)}}\left(y_{1: T}\right)$ and $ G_{T}^{'}\left(\theta^{(0)}\right)$.\\

\For{i = 1,..,L}{
Set $r_{i} \leftarrow r_{i-1} - \frac{\epsilon}{2}$ $ G_{T}^{'}\left(\theta^{(i-1)}\right)$.\\

Set $\theta^{(i)}  \leftarrow \theta^{(i-1)} + \epsilon r_{i}$.\\

Run an SMC algorithm, using $N$ particles, targeting $p_{\theta^{(i)}}\left(h_{1: T} \mid y_{1: T}\right)$ then store $ G_{T}^{'}\left(\theta^{(i)}\right)$ and also store a sample $H_{1: T}^{*}$ and the marginal likelihood estimate $\hat{p}_{\theta^{(L)}}\left(y_{1: T}\right)$ if $i=L$.\\

Set $r_{i} \leftarrow r_{i} - \frac{\epsilon}{2}$ $G_{T}^{'}\left(\theta^{(i)}\right)$. \\
}

With probability
  \\
$
 \min\left(1, \exp\left(\hat{Ha}(\theta^{(0)},r_{0}) - \hat{Ha}(\theta^{(L)},r_{L})\right)\right)
$, 
  \\
Set $\left(\theta^{(k)},r^{(k)}  \right) = \left(\theta^{(L)},-r_{L} \right)$ and $H_{1: T}^{(k)} = H_{1: T}^{*}$; Otherwise set $\theta^{(k)} = \theta^{(k-1)}$ and $H_{1: T}^{(k)} = H_{1: T}^{(k-1)}$.

  }

\caption{Particle Hamiltonian Monte Carlo}
\label{phmc_algorithm}
\end{algorithm}

\subsection{Discussion}
\begin{flushleft}
Importantly, PHMC does not require estimates of the gradients of the latent variables' posterior. This aspect is crucial for models with complex dynamics, where estimating latent variable gradients could become prohibitively expensive. 
While it is true that replacing, in each $i$-th iteration, $ G_{T}^{'}\left(\theta^{(i-1)}\right)$ and $ G_{T}^{'}\left(\theta^{(i)}\right)$ with $ G_{T}^{}\left(\theta^{(i)}\right)$ and $ G_{T}^{}\left(\theta^{(i-1)}\right)$ respectively could significantly reduce computational time, we believe this would deteriorate the performance of the algorithm, especially without increasing the number of particles \( N \). This is because using the estimation approach that gives $G_{T}^{}\left(\theta^{(i)}\right)$ and $ G_{T}^{}\left(\theta^{(i-1)}\right)$, may suffer from path degeneracy, leading to gradient estimates with high variability. 

There are three key parameters that the user must choose: the number of particles \( N \), the number of leapfrog steps \( L \), and the step size \( \epsilon \).  For the number of particles \( N \), it is well-known that increasing \( N \) generally improves accuracy, but at the cost of greater computational expense. Ideally, \( N \) should be at least larger than the time horizon \( T \). We recommend initially measuring the variance of gradient estimates over several runs of the SMC algorithm with different values of \( N \), then selecting the number of particles that results in reasonably low variance in both the gradient and likelihood estimates.

Tuning the other two parameters, \( L \) and \( \epsilon \), for the Hamiltonian Monte Carlo (HMC) component can be more challenging, and this task is beyond the scope of our work. Briefly, For a fixed $L$, an $\epsilon$ that is too large will lead to low acceptance rates, while an $\epsilon$ that is too small will waste computational resources by taking numerous small steps. For a fixed $\epsilon$, if $L$ is too small, successive samples will be closely spaced, causing undesirable random walk behavior and slow mixing. Conversely, if $L$ is too large, the computational cost would be increased and HMC will produce trajectories that loop back and retrace their paths \parencite{nuts}. 
Furthermore, by setting $L=1$, HMC simplifies to the Metropolis-Adjusted Langevin Algorithm (MALA); in our context, this results in particle MALA, as introduced in \textcite{dahlin2013particle} and \textcite{nemeth2016particle}. In \textcite{nemeth2016particle}, they adopted an alternative gradient estimation method proposed in \textcite{nemeth2013particle}, which has a linear computational cost. This method uses kernel density estimation with Gaussian kernels to approximate gradients, followed by Rao-Blackwellization. Additionally, \parencite{corenflos2024particle} introduced particle MCMC variants of both MALA and auxiliary MALA (aMALA)\parencite{titsias2018auxiliary}, where they used the method suggested by \textcite{nemeth2013particle}. 
We suggest that this gradient estimation technique could potentially serve as a viable alternative to \textcite{poy} in the context of particle Hamiltonian Monte Carlo (HMC), primarily due to its lower computational cost. In many cases, the reduced complexity of this approach can make it more efficient based on the numerical studies in \textcite{nemeth2013particle}. However, the main challenge arises in scenarios where the score vector estimation, based on Gaussian kernels may be very inaccurate. Moreover, there is also another fast alternative that can be used in this context, which was proposed by \textcite{Klaas2006}.
To further enhance the efficiency of the algorithm, another recommendation is to employ a Sequential Monte Carlo (SMC) algorithm with a better proposal distribution, the SMC algorithm can provide estimates of the marginal likelihood with smaller variability. This, in turn, would help to significantly reduce the estimation variance in the acceptance probability during the Metropolis-Hastings steps.
\end{flushleft}

\section{Numerical experiments}
\begin{flushleft}
    In this section, we evaluate the proposed method's effectiveness using data simulated from two models in addition to an application based on real data. Refer to the Appendix for the gradient calculations of the probability densities with respect to the parameters in the two state-space models considered.
\end{flushleft}

\subsection{Poisson count model}
\label{poissonhmcsubsection}
\begin{flushleft}
    Here, we considered the following state-space model that is used for modelling count data:
\end{flushleft}

\begin{equation} 
\label{model2}
 \begin{split}
     Y_{t}|H_{t} = h_{t}   &\sim Po(e^{h_{t}+\alpha}),  \\
     H_{t} \mid H_{t-1} = h_{t-1}  &\sim \mathcal{N}\left(\rho h_{t-1}, \sigma_h^{2} \right), \\
      H_{1} &\sim \mathcal{N}\left(0, \frac{\sigma_h^{2}}{1- \rho^{2}} \right),
 \end{split}
\end{equation}

\begin{flushleft}
where $t \in 1:T$ and $y_{t}$ is an observation of $Y_{t}$. We simulate data from this model using $T = 100$, $\sigma_h = 0.2$, $\alpha = 0.5$ and $\rho=0.8$. 
\end{flushleft}

\begin{figure}[h]
 \begin{subfigure}[]{0.6\textwidth}
         \centering
         \includegraphics[width=\linewidth]{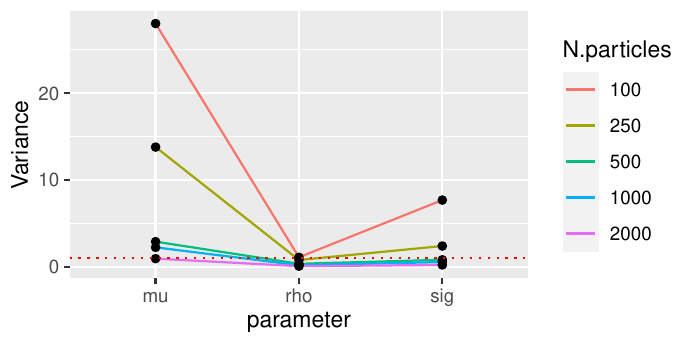}
         \caption{}
     \end{subfigure}
      \begin{subfigure}[]{0.4\textwidth}
         \centering
         \includegraphics[width=\linewidth]{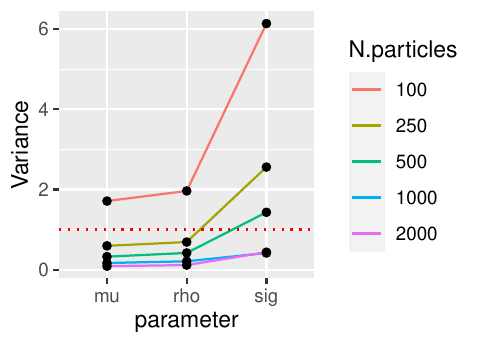}
         \caption{}
     \end{subfigure}
          
	    	\caption{The variance of the estimates of the gradients related to the three parameters obtained by the SMC algorithm related to  (a) $S_{T}(\theta)$ and  (b) $S_{T}^{'}(\theta)$ with different values of the number of particles $N$=[100,250,500,1000,2000] for $T=100$. The dotted red line represents the value 1.}
      \label{fig:p2}
\end{figure}

\begin{flushleft}
 First, we intend to compare the estimates $S_{T}(\theta)$ and $S_{T}^{'}(\theta)$ based on their variability. To do this, we run 10 particle filters on the simulated data, each time using a different number of particles 
$N$ (with 
values specified in [100,250,500,1000,2000]). We then calculate the variance of the estimates for each instance.
 Figure \ref{fig:p2} illustrates how the particle count influences the variance of gradient estimates  in both methods of obtaining estimates of the gradients. Throughout all experiments, the parameters were fixed at $\sigma_h = 0.8,\rho=0$ and $\alpha=1 $  when running the SMC algorithms. As the number of particle decreases, the variance of the estimates grows which affects the performance of pseudo-marginal MCMC. The second method is observed to yield significantly improved gradient estimates in terms of variance. It is also worth noting that the performance differs from one parameter to another as they have different posterior densities. In their experiments, \textcite{conc1} used \(N^2\) particles for the \(O(N)\) method and \(N\) particles for the \(O(N^2)\) method, ensuring a fair comparison in terms of computational cost. The results demonstrate that the \(O(N)\) method achieves better performance with respect to bias, whereas the \(O(N^2)\) method yields lower variance. Moreover, the mean squared error (MSE) of the rescaled estimates increases linearly in both cases, aligning with the theoretical properties of \textcite{hmcexperiment1,hmcexperiment2,hmcexperiment3}.
\end{flushleft}

\begin{figure}[htb]
\centering
\includegraphics[width=13cm]{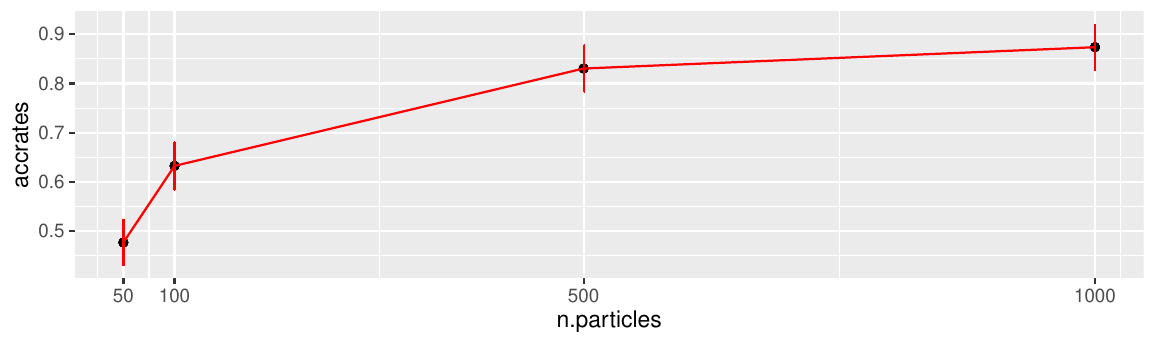}
\caption{The acceptance rates of 10 runs of particle HMC based on a simulated data set with 100 parameters vs the number of particles $N$=[50,100,500,1000]. The dark dots refer to the median of the rates while the intervals represent the median +/- the standard deviation of the rates. The time horizon $T$ of all simulated data sets was set to 100. For all experiments, 5000 iterations were used, the burn-in was set to 1000, L = 5, $\epsilon$ = 0.05 and the thinning to 10.}
\label{fig:p3}
\end{figure}

\begin{flushleft}
    For each $N$ in the set [50,100,500,1000], we ran 10 Markov chains of particle Hamiltonian Monte Carlo. Each chain has 5000 iterations, a burn-in period of 1000 iterations, and a thinning interval of 10. The number of the leapfrog steps $L$ was fixed at 5 whereas the step size $\epsilon$ was set to 0.05. We opted for these prior distributions for the parameters $\rho, \alpha, \sigma_h$, which are assumed to be unknown: $\rho \sim Uniform[-1,1], \alpha \sim \mathcal{N}(m_{\alpha},s_{\alpha}^{2}) $ and $\frac{1}{\sigma_h^{2}} \sim Gamma(a,b)$, and we set a = b = 0.01,  $m_{\alpha}$ = 0 and $s_{\alpha} = 10$. The initial values for the parameters, in the Markov chains, are sampled from their prior distributions. Figure \ref{fig:p3} displays the acceptance rates of these chains in relation to the number of particles used. The dark dots indicate the median acceptance rates, while the intervals represent the median plus or minus the standard deviation of the rates.

    One can observe a positive correlation, on average, between the acceptance rate and the number of particles. This correlation may be attributed to the phenomenon where employing a limited number of particles leads to noisy estimates of the gradients, consequently impacting the efficacy of particle Hamiltonian Monte Carlo (pHMC). Notably, the acceptance rates tend to stabilize and show minimal change when a sufficiently large number of particles is utilized. 
\end{flushleft}

\begin{figure}[h]
\centering
\includegraphics[width=15cm]{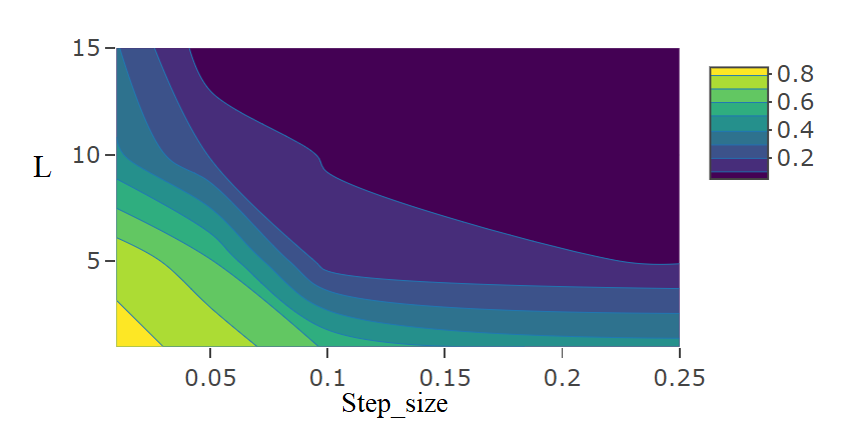}
\caption{The median of acceptance rates of 10 runs of particle HMC (with 5000 iterations, the burn-in was set to 1000 and the thinning to 10) on data simulated from the Poisson state-space model, ranging from low (dark blue) to high (yellow), as a function of the step size $\epsilon$ (x-axis) and the number of leapfrog steps $L$ (y-axis).}
\label{fig:p4}
\end{figure}

\begin{flushleft}
   In another experiment, we investigated the effect of the step size $\epsilon$ and the number of the leapfrog steps $L$ on the performance of the particle Hamiltonian Monte Carlo in terms of the acceptance rate.  The effectiveness of HMC heavily relies on selecting appropriate values for $\epsilon$ and $L$. We adjust the step size $\epsilon$ within the range [0.005, 0.25] and 
$L$ within [1, 15]. For each configuration, we run 10 particle HMC Markov chains with 5000 iterations, using a burn-in period of 1000 and a thinning parameter of 10. The particle filter algorithms are configured with 400 particles. Similar to the previous experiment,  the starting points are sampled from their prior distributions. Figure \ref{fig:p4} presents the median acceptance rates in relation to both $L$ (y-axis) and $\epsilon$ (x-axis). Based on the simulated data, it is observed that higher values of the step size or the number of leapfrog steps result in lower acceptance rates. The highest acceptance rates occur when both $\epsilon$  and $L$ are low.

\end{flushleft}

\subsection{ Linear Gaussian State-Space Model with an average shift term}
\begin{flushleft}
    To illustrate the approach that was proposed we consider data simulated from a one-dimensional linear gaussian state space model where there is an average of $d$ parameters that are the elements of $\boldsymbol{\kappa} = \left(\kappa_{1},..,\kappa_{d} \right)$:
\begin{equation} 
\label{model1}
 \begin{split}
     Y_{t}|H_{t} = h_{t}   &\sim \mathcal{N}\left(h_{t} , \sigma^{2}_{y} \right),  \\
     H_{t} \mid H_{t-1} = h_{t-1}  &\sim \mathcal{N}\left(\rho h_{t-1} + \frac{1}{d}\sum_{j=1}^{d} \kappa_{j}, \sigma^{2}_{h} \right), \\
      H_{1} &\sim \mathcal{N}\left(0, \frac{\sigma^{2}_{h}}{1- \rho^{2}} \right), 
 \end{split}
\end{equation}
\end{flushleft}
\begin{flushleft}
where $t \in 1:T$ and $y_{t}$ is an observation of $Y_{t}$. For each $d \in$ [5,25,50,75,100], we simulate data from this model using $T = 100$, $\sigma_h = 0.2$, $\sigma_y = 0.25$ and $\rho=0.8$. The values of $\kappa_i$ for $i \in 1:d$ were selected so that their average was equal to 0.5. In this case, the number of parameters, $ d_{\theta},$ is $d+3$ as $\theta= \left(\boldsymbol{\kappa}, \sigma_y, \sigma_h, \rho \right)$ but note that $Y_t$ and $H_t$ are both one-dimensional for all time steps $t \leq T$. We run 10 Markov chains of both particle Metropolis-Hastings with random walk and particle Hamiltonian Monte Carlo on that simulated data. For each chain, 5000 iterations were considered, the burn-in period was 1000 and a thinning
parameter of 10 was also considered. The number of particles in the particle filter algorithms was 500. The values of the first iteration were sampled from their prior distributions. The main motivation behind opting for this model is to assess how the proposed method performs in case of dealing with a high-dimensional parameter space.
\end{flushleft}

\begin{figure}[h]
         \centering
         \includegraphics[width=1\linewidth]{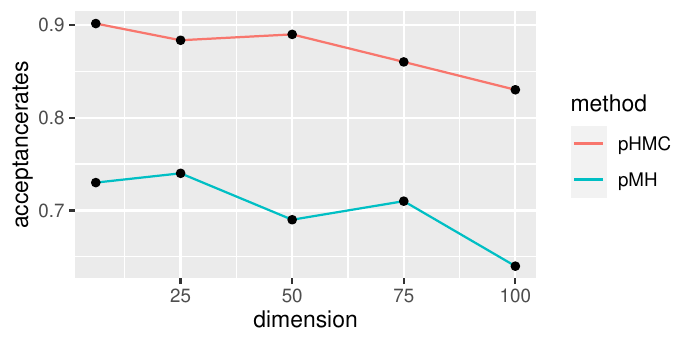}  
	    	\caption{ The median of the acceptance rates of 10 runs of both the particle HMC (red) and the particle marginal Metropolis-Hastings with random walk (blue) vs the total number of parameters used in the model.}
      \label{fig:p1}
\end{figure}

\begin{flushleft}
    \textcite{neal2012mcmc}, building on \parencite{creutz1988global}, offers an informal rationale for scaling $\epsilon$ as $d_{\theta}^{-1/4}$, particularly for targets that decompose into products of d independent components. Consequently, to keep the integration time $L\epsilon$ constant, $L$ should be scaled as $d_{\theta}^{1/4}$. Hence, for each $d_{\theta}$ we set $\epsilon = 0.025 d_{\theta}^{-1/4}$ for both particle Hamiltonian Monte Carlo and particle Metropolis-Hastings with random walk proposal. In the case of pHMC, the number of the leapfrog steps was set to $L = 5 d_{\theta}^{1/4}$.\\  Figure \ref{fig:p1} illustrates the median acceptance rates from 10 runs of both the particle HMC (in red) and the particle marginal Metropolis-Hastings with random walk (in blue) against the total number of parameters used in the model. The results show that particle HMC achieves higher acceptance rates compared to the other method. Additionally, there is a noticeable dependence on the number of dimensions, which is slightly more evident in the case of particle marginal Metropolis-Hastings with random walk i.e. the median of acceptance rates is lower when the number of parameters is large.  
\end{flushleft}

\subsection{Examining the underlying temporal patterns in Strava data}
\begin{flushleft}
The advent of crowdsourced applications has transformed methods of data collection and analysis. Strava, a widely-used fitness app, exemplifies how user-generated data can provide valuable insights into physical activity patterns so extracting hidden temporal structures and patterns from such data is interesting. We will illustrate how the application of the particle HMC method discussed in Section \ref{phmcsubsection} can be applied in this context. We will illustrate this method on the number of crowd-sourced cycling activities in the thoroughfare of Broomielaw in the city of Glasgow. The length of the time series is 48 hours (based on 2 days of hourly counts in September 2013). As we are dealing with count data, the standard choice is a Poisson state-space model so we opted for the same model used in the experiments with simulated count data in subsection \ref{poissonhmcsubsection}, 
\end{flushleft}

\begin{equation} 
 \begin{split}
     Y_{t}|H_{t} = h_{t}   &\sim Po(e^{h_{t}+\alpha})  \\
     H_{t} \mid H_{t-1} = h_{t-1}  &\sim \mathcal{N}\left(\rho h_{t-1}, \sigma_h^{2} \right) \\
      H_{1} &\sim \mathcal{N}\left(0, \frac{\sigma_h^{2}}{1- \rho^{2}} \right),
 \end{split}
\end{equation}

\begin{flushleft}
   where $t \in 1:48$, $Y_{t}$ is the number of bike rides in Broomielaw at hour $t$ and the $H_{t}$ latent variable evolves according to a Gaussian random walk and the observations are modelled as Poisson distribution with mean $exp(h_{t}+ \alpha)$ . To implement the particle Hamiltonian Monte Carlo, we used Integrated Nested Laplace Approximation to initialise the parameters using values that are close to the posterior modes of $\alpha, \rho$ and $\sigma_h$ as it was done in \textcite{amri} to reduce the computational cost. We used 100 particles in the particle filter to approximate the marginal likelihood. The number of leapfrog steps \(L\) was set to 10, and the step size \(\epsilon\) was chosen as 0.0015, resulting in a relatively small integration time of 0.015. The Markov chain ran for 10,000 iterations, with the first 1,000 iterations discarded as burn-in. Additionally, we applied thinning by retaining every 10th iteration to minimize autocorrelation. The acceptance rate was 0.532.
\end{flushleft}

\begin{figure}[htbp]	
	\centering
	\includegraphics[width=.95\linewidth]{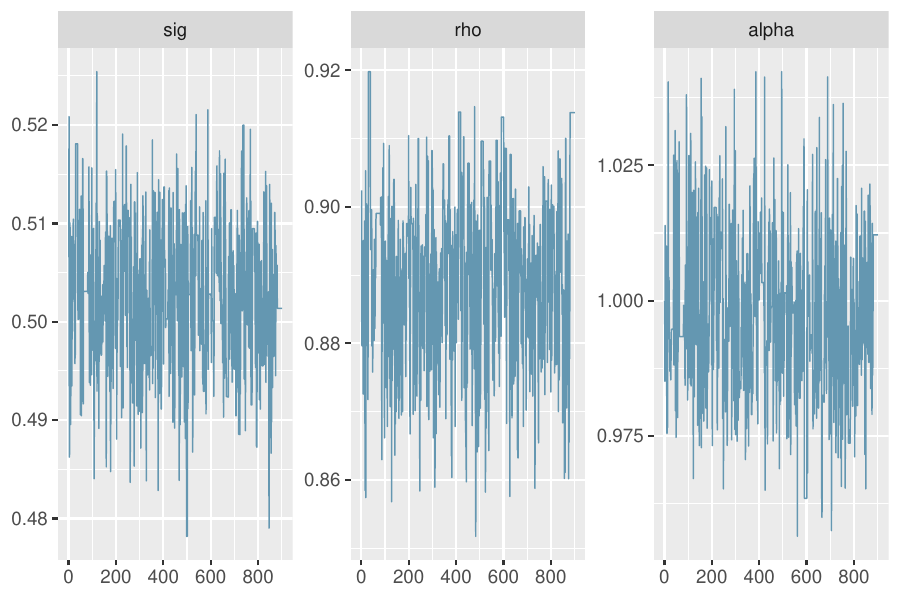}
	\caption{Trace plots related to $\sigma$ (left), $\rho$ (middle) and $\alpha$ (right).}\label{Fig:hmcapp1}

\end{figure}

\begin{figure}[htbp]	
	\centering
	\includegraphics[width=.8\linewidth,height=5cm]{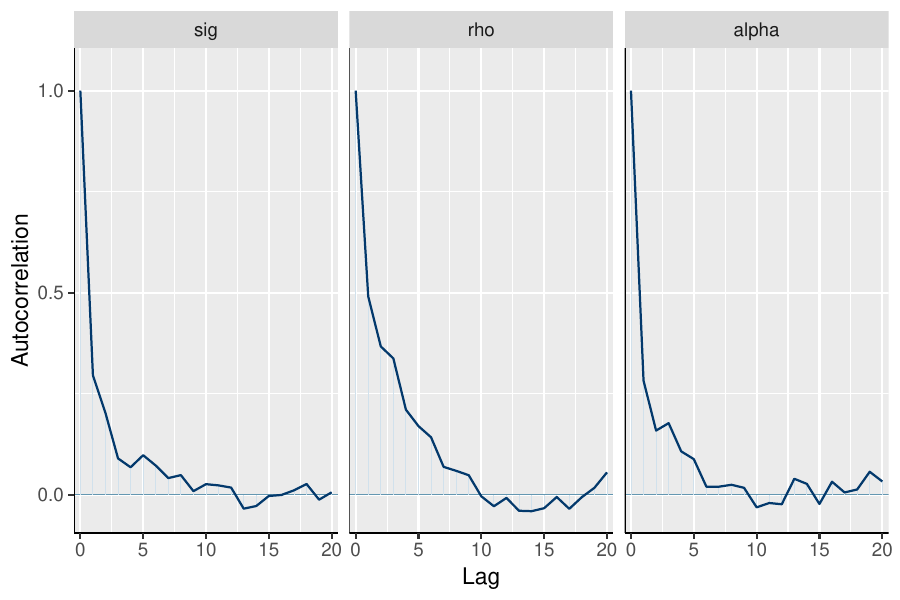}
	\caption{Autocorrelation function  plots related to $\sigma$ (left), $\rho$ (middle) and $\alpha$ (right).}\label{Fig:hmcapp3}

\end{figure}

\begin{figure}[htbp]	
	\centering
	\includegraphics[width=.8\linewidth,height=5cm]{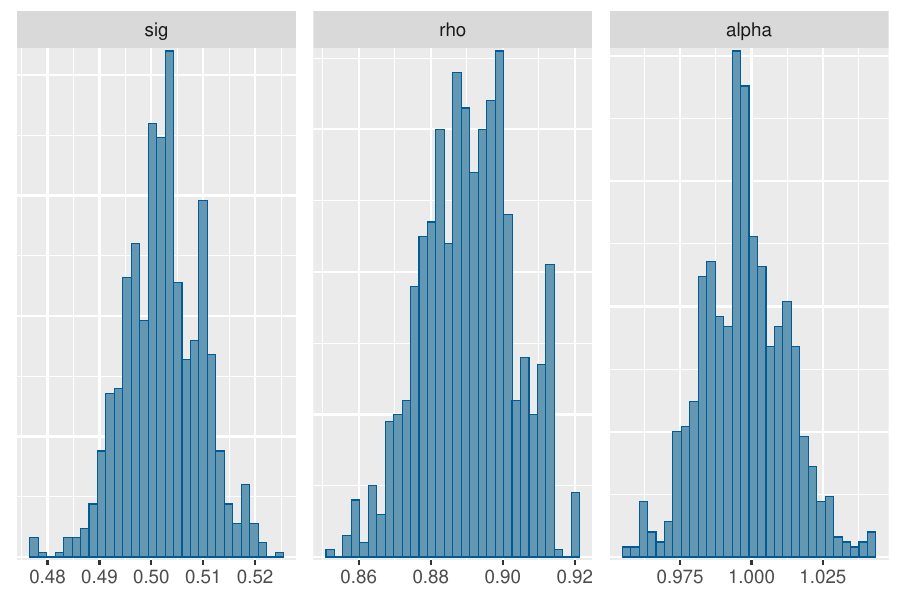}
	\caption{Histograms of the particle HMC draws related to $\sigma$ (left), $\rho$ (middle) and $\alpha$ (right).}\label{Fig:hmcapp2}

\end{figure}
\begin{figure}[htbp]	
	\centering
	\includegraphics[width=.95\linewidth]{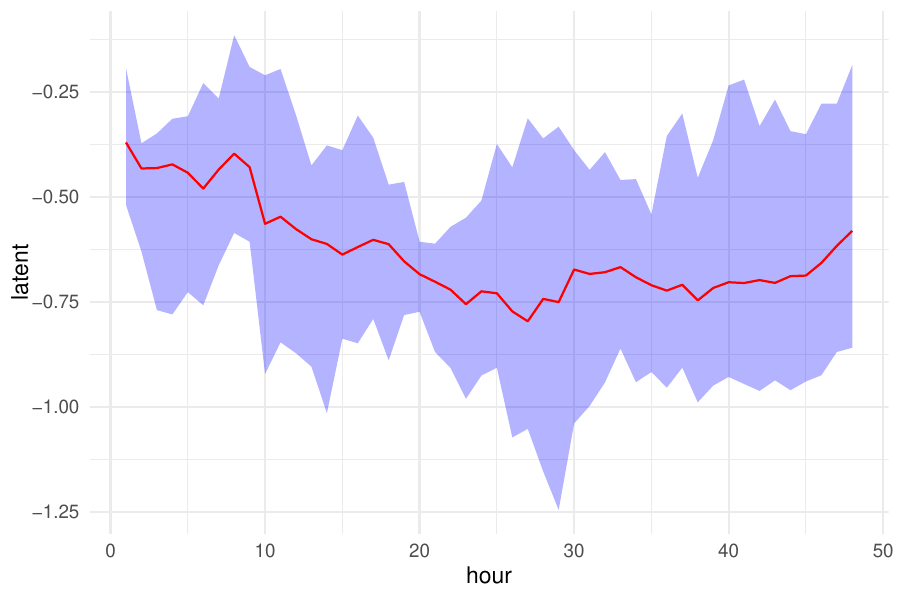}
	\caption{The
posterior mean (in red) of the particles of the latent variables (obtained using particle Hamiltonian Monte Carlo) is plotted against time (in years) and the 95\% credible intervals are shown as shaded regions
around the posterior mean.}\label{Fig:hmcapp6}
\end{figure}

\begin{flushleft}
    Figures \ref{Fig:hmcapp1} and \ref{Fig:hmcapp3} show the trace plots and the autocorrelation function (ACF) plots related to the model parameters, respectively.
    The histograms of the posterior samples are illustrated in \ref{Fig:hmcapp2} and they display Gaussian-like shapes. Based on Figures \ref{Fig:hmcapp2} and \ref{Fig:hmcapp3}, the posterior samples of $\rho$ seem to have higher autocorrelation and kurtosis when compared to the other two parameters whose samples have autocorrelations that decline rapidly within a few lags. Figure \ref{Fig:hmcapp6} displays the posterior mean estimate (in red) of the latent variable particles (sampled using particle Hamiltonian Monte Carlo), plotted against time (in years). The $95\%$  credible intervals are represented as shaded regions around the posterior mean. The samples appear to be negative, with a decreasing trend in the posterior mean on the first day, though it fluctuates slightly during the second day. Regarding the confidence intervals, those associated with the latent variables on the second day are generally wider than the corresponding intervals on the first day.
    
\end{flushleft}

\section{Conclusions}
The proposed method, namely particle Hamiltonian Monte Carlo, demonstrated superior performance compared to Particle Marginal Metropolis-Hastings (PMMH) using a Gaussian random walk proposal, especially in scenarios involving models with a large parameter space. However, a key limitation emerged with regard to computational efficiency as the computational cost of gradient estimation by \parencite{poy} is \( O(N^2) \), which can make the approach prohibitive in terms of runtime.  To address this, one possible direction for future research would be to implement a strategy of adaptively choosing the number of particles, potentially borrowing from methods proposed in \textcite{adaptivenparticles1,adaptivenparticles2}. This adaptive strategy could dynamically adjust the number of particles based on the needs of the model, potentially enhancing computational efficiency without sacrificing accuracy. \\

Additionally, while this work did not cover automatically tuning the the step size \( \epsilon \) and the number of leapfrog steps, future work could investigate this area to optimize the algorithm’s performance further. An interesting direction here could involve developing a particle MCMC adaptation of the No-U-Turn Sampler (NUTS) \parencite{nuts}, known for its automated tuning, or exploring a particle-based randomized Hamiltonian Monte Carlo (HMC), building on \parencite{bou2017}. This would allow for more efficient exploration of complex posterior distributions of parameters in state-space models. Furthermore, one can potentially consider a state-dependent mass matrix  in particle Hamiltonian Monte Carlo instead of the identity matrix like in the method Riemannian manifold HMC \parencite[]{RMHMC}. Another promising idea for future exploration could involve creating a particle-based variant of the MALT algorithm \parencite{riou2022} and comparing it with our proposed framework. This comparison could yield insights into the relative strengths and limitations of these methods. \newline


\section*{Appendix}
\label{appendix}
\subsection*{Proof of the Fisher's identity}
We start by writing the likelihood as 
$$
p_\theta\left(y_{1:T}\right)=\int p_\theta\left(h_{1:T}, y_{1:T}\right) \mathrm{d} h_{1:T}
$$
Then note that we can write $\nabla_\theta \log p_\theta\left(y_{1:T}\right)$ in the following way by differentiating the logarithm.
$$
\nabla_\theta \log p_\theta\left(y_{1:T}\right)=\frac{\nabla_\theta p_\theta\left(y_{1:T}\right)}{p_\theta\left(y_{1:T}\right)} = \frac{\nabla_\theta \int p_\theta\left(h_{1:T}, y_{1:T}\right) \mathrm{d} h_{1:T}}{p_\theta\left(y_{1:T}\right)}
$$
 Assuming that all functions are regular enough to perform change of integration and differentiation then we get

$$
\nabla_\theta \log p_\theta\left(y_{1:T}\right)=\int \frac{\nabla_\theta p_\theta\left(h_{1:T}, y_{1:T}\right)}{p_\theta\left(y_{1:T}\right)} \mathrm{d} h_{1:T}
$$

Given that,
$$
\nabla_\theta p_\theta\left(h_{1:T}, y_{1:T}\right)=\nabla_\theta \log p_\theta\left(h_{1:T}, y_{1:T}\right) p_\theta\left(h_{1:T}, y_{1:T}\right)
$$

We can express  $\nabla_\theta \log p_\theta\left(y_{1:T}\right)$ in the following way,
\begin{equation*}
\begin{aligned}
\nabla_\theta \log p_\theta\left(y_{1:T}\right)&=\int \nabla_\theta \log p_\theta\left(h_{1:T}, y_{1:T}\right) \frac{p_\theta\left(h_{1:T}, y_{1:T}\right)}{p_\theta\left(y_{1:T}\right)} \mathrm{d} h_{1:T} \\
&=\int \nabla_\theta \log p_\theta\left(h_{1:T}, y_{1:T}\right) p_\theta\left(h_{1:T} \mid y_{1:T}\right) \mathrm{d} h_{1:T} 
\end{aligned}
\end{equation*}

\subsection*{Derivatives of the probability densities in the state-space models with respect to the parameters}

\begin{flushleft}
    I) Model 1:
\end{flushleft}

\begin{flushleft}
    The logarithm of the observation density at time $t$ is given as 
    \begin{equation*}
 \log p\left(y_t \mid h_t\right)=y_t\left(h_t+\alpha\right)-\exp \left(h_t+\alpha\right)-\log \left(y_t\,!\right).
\end{equation*}
\end{flushleft}

\begin{flushleft}
    Hence, the derivatives of the logarithm of the density above, with respect to the parameters are the following:
\begin{equation*}
    \frac{\partial \log p\left(y_t \mid h_t\right)}{\partial \alpha}=y_t-\exp \left(h_t+\alpha\right),
\frac{\partial \log p\left(y_t \mid h_t\right)}{\partial \rho} = \frac{\partial \log p\left(y_t \mid h_t\right)}{\partial \sigma_h} = 0.
\end{equation*}
\end{flushleft}

\begin{flushleft}
    The logarithm of the transition density from time $t-1$ to $t$ is the following
    \begin{equation*}
    \log p\left(h_t \mid h_{t-1}\right)=-\log \sigma_h-\log \sqrt{2 \pi}-\frac{1}{2}\left(\frac{h_t- \rho h_{t-1}}{\sigma_h}\right)^2.
\end{equation*}
\end{flushleft}

\begin{flushleft}
    Hence, the derivatives of the logarithm of the density above, with respect to the parameters are the following:
\begin{equation*}
    \begin{aligned} 
  &  \frac{\partial \log p\left(h_t \mid h_{t-1}\right)}{\partial \rho}  = \frac{1}{\sigma_h^{2}} \left(h_t h_{t-1} - \rho h_{t-1}^{2}\right),
    \frac{\partial \log p\left(h_t \mid h_{t-1}\right)}{\partial \sigma_h}  = \frac{-1}{\sigma_h} 
 +\sigma_h^{-3}\left(h_t-\rho h_{t-1}\right)^2, \\
 &\frac{\partial \log p\left(h_t \mid h_{t-1}\right)}{\partial \alpha}= 0.
\end{aligned} 
\end{equation*}
\end{flushleft}

\begin{flushleft}
    The logarithm of the density of the first latent variable is given below 
    \begin{equation*}  
    \log p\left(h_1\right)=-\log \sigma_h+\frac{1}{2} \log \left(1-\rho^2\right)-\log \sqrt{2 \pi}-\frac{1}{2} h_1^2 \times \frac{\left(1-\rho^2\right)}{\sigma_h^2}.
\end{equation*} 
\end{flushleft}

\begin{flushleft}
    Hence, the derivatives of the logarithm of the density above, with respect to the parameters are the following:
\begin{equation*}
    \frac{\partial \log p\left(h_1\right)}{\partial \rho}=\frac{h_1^2 \rho}{\sigma_h^2}-\frac{\rho}{1-\rho^2}, \quad \frac{\partial \log \rho\left(n_n\right)}{\partial \sigma_h}=\frac{-1}{\sigma_h}+\sigma_h^{-3} h_1^2\left(1-\rho^2\right) , \frac{\partial \log p\left(h_1\right)}{\partial \alpha}=0.
\end{equation*}
\end{flushleft}

\begin{flushleft}
    II) Model 2:
\end{flushleft}

\begin{flushleft}
    The logarithm of the observation density at time $t$ is given as 
    \begin{equation*}
    \log p\left(y_t \mid h_t\right)=-\log \sigma_y-\log \sqrt{2 \pi}-\frac{1}{2}\left(\frac{y_t-h_t}{\sigma_y^2}\right)^2. 
\end{equation*}
\end{flushleft}

\begin{flushleft}
    Hence, the derivatives of the logarithm of the density above, with respect to the parameters are the following:
\begin{equation*}
 \frac{\partial \log p\left(y_t \mid h_t\right)}{\partial \sigma_y}=-\frac{1}{\sigma_y}+\sigma_y^{-3}\left(y_t-h_t\right)^2, \frac{\partial \log p\left(y_t \mid h_t\right)}{\partial \rho} = \frac{\partial \log p\left(y_t \mid h_t\right)}{\partial \sigma_h} = \frac{\partial \log p\left(y_t \mid h_t\right)}{\partial \kappa_{i}} = 0.
\end{equation*}
\end{flushleft}

\begin{flushleft}
    The logarithm of the transition density from time $t-1$ to $t$ is the following
\end{flushleft}
\begin{equation*}
   \log p\left(h_t \mid h_{t-1}\right)=-\log \sigma_h-\log \sqrt{2 \pi}-\frac{1}{2}\left(\frac{h_t- \frac{1}{d}\sum_{j=1}^{d} \kappa_{j} -\rho h_{t-1}}{\sigma_h}\right)^2. 
\end{equation*}

\begin{flushleft}
    Hence, the derivatives of the logarithm of the density above, with respect to the parameters are the following:
\end{flushleft}
\begin{equation*}
\begin{aligned}
 &   \frac{\partial \log p\left(h_t \mid h_{t-1}\right)}{\partial \rho}  = \frac{h_{t-1}}{\sigma_h^2}\left(h_t-\frac{1}{d}\sum_{j=1}^{d} \kappa_{j}-\rho h_{t-1}\right), \\
& \frac{\partial \log p\left(h_t \mid h_{t-1}\right)}{\partial \kappa_{i}}  = \frac{1}{ 
 \sigma_h^2 d}\left(h_t-\frac{1}{d} \kappa_{i} - \frac{1}{d}\sum_{j \neq i}^{d} \kappa_{j}-\rho h_{t-1}\right), \\
&    \frac{\partial \log p\left(h_t \mid h_{t-1}\right)}{\partial \sigma_h}  = \frac{-1}{\sigma_h} 
 +\sigma_h^{-3}\left(h_t-\frac{1}{d}\sum_{j=1}^{d} \kappa_{j}-\rho h_{t-1}\right)^2, \frac{\partial \log p\left(h_t \mid h_{t-1}\right)}{\partial \sigma_y}= 0.
\end{aligned} 
\end{equation*}
 
\begin{flushleft}
    The logarithm of the density of the first latent variable is given below 
    \begin{equation*}
    \log p\left(h_1\right)=-\log \sigma_h+\frac{1}{2} \log \left(1-\rho^2\right)-\log \sqrt{2 \pi}-\frac{1}{2} h_1^2 \times \frac{\left(1-\rho^2\right)}{\sigma_h^2}.
\end{equation*}
\end{flushleft}

\begin{flushleft}
    Hence, the derivatives of the logarithm of the density above, with respect to the parameters are the following:
\end{flushleft}
\begin{equation*}
\begin{aligned}
& \frac{\partial \log p\left(h_1\right)}{\partial \rho}=\frac{h_1^2 \rho}{\sigma_h^2}-\frac{\rho}{1-\rho^2}, \quad \frac{\partial \log \rho\left(n_n\right)}{\partial \sigma_h}=\frac{-1}{\sigma_h}+\sigma_h^{-3} h_1^2\left(1-\rho^2\right) , \frac{\partial \log p\left(h_1\right)}{\partial \sigma_y}= \frac{\partial \log p\left(h_1\right)}{\partial \kappa_{i}} = 0 . \\
\end{aligned}
\end{equation*}
\subsection*{Application}
\begin{figure}[htbp]	
	\centering
	\includegraphics[width=\linewidth,height=6cm]{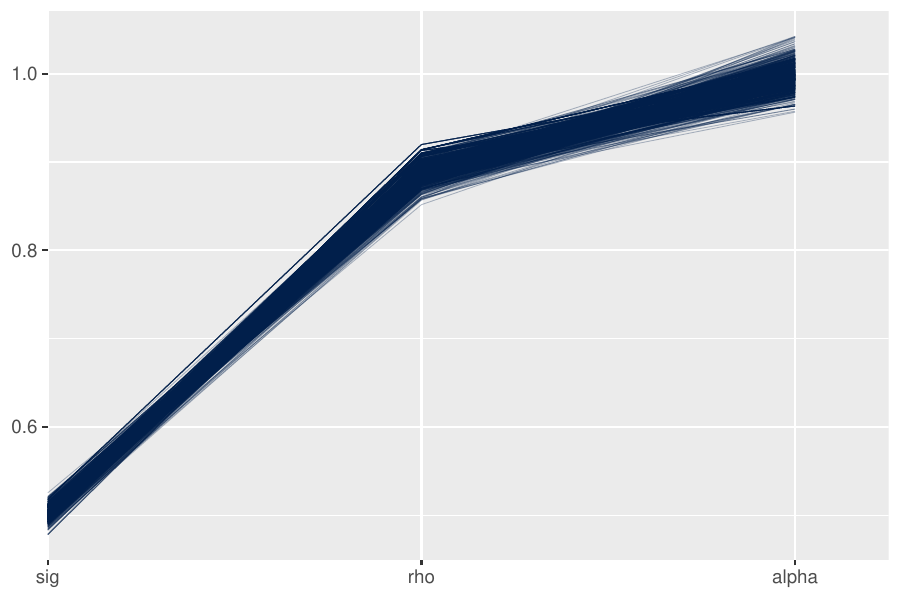}
	\caption{Parallel coordinates plot of the particle HMC draws related to $\sigma$ (left), $\rho$ (middle) and $\alpha$ (right).}\label{Fig:hmcapp4}

\end{figure}

\section*{Acknowledgements}
\begin{flushleft}
     The first author would like to thank Adam M. Johansen, Nikolas Kantas and Finn Lindgren for helpful comments and he is supported by Edinburgh Future Cities studentship.
\end{flushleft}

\printbibliography

@article{RMHMC,
  author    = {Girolami, M. and Calderhead, B.},
  title     = {Riemann Manifold Langevin and Hamiltonian Monte Carlo Methods},
  year      = {2011},
  journal   = {Journal of the Royal Statistical Society: Series B (Statistical Methodology)},
  volume    = {73},
  number    = {2},
  pages     = {123--214}
}

@article{roberts1996exponential,
  title={Exponential convergence of Langevin distributions and their discrete approximations},
  author={Roberts, G. O. and Tweedie, R. L.},
  journal={Bernoulli},
  volume={2},
  pages={341--363},
  year={1996}
}

@ARTICLE{amri,
  title         = "Designing proposal distributions for particle filters using
                   integrated Nested Laplace Approximation",
  author        = "Amri, A.",
  month         =  may,
  year          =  2023,
  copyright     = "http://arxiv.org/licenses/nonexclusive-distrib/1.0/",
  archivePrefix = "arXiv",
  primaryClass  = "stat.CO",
  eprint        = "2305.03552"
}

@article{adaptivenparticles1,
  author    = {Elvira, V. and Míguez, J. and Djurić, P. M.},
  title     = {Adapting the Number of Particles in Sequential Monte Carlo Methods Through an Online Scheme for Convergence Assessment},
  year      = {2016},
  journal   = {IEEE Transactions on Signal Processing},
  volume    = {65},
  number    = {7},
  pages     = {1781--1794}
}

@article{adaptivenparticles2,
  author    = {Elvira, V. and Míguez, J. and Djurić, P. M.},
  title     = {On the Performance of Particle Filters with Adaptive Number of Particles},
  year      = {2021},
  journal   = {Statistics and Computing},
  volume    = {31},
  pages     = {1--18}
}

@article{riou2022,
  author    = {Riou-Durand, L. and Vogrinc, J.},
  title     = {Metropolis Adjusted Langevin Trajectories: A Robust Alternative to Hamiltonian Monte Carlo},
  year      = {2022},
  journal   = {arXiv preprint},
  eprint    = {arXiv:2202.13230}
}

@article{bou2017,
  author    = {Bou-Rabee, N. and Sanz-Serna, J. M.},
  title     = {Randomized Hamiltonian Monte Carlo},
  year      = {2017},
  journal   = {Annals of Applied Probability}
}

@book{int1,
  author = {O. Capp{\'e} and E. Moulines and T. Ryd{\'e}n},
  title = {Inference in Hidden Markov Models},
  publisher = {Springer},
  year = {2005}
}

@book{Chopin2020,
  author = {N. Chopin and O. Papaspiliopoulos},
  title = {An Introduction to Sequential Monte Carlo},
  publisher = {Springer International Publishing},
  year = {2020}
}

@incollection{hairer2012numerical,
  title={Numerical solution of ordinary differential equations},
  author={Hairer, E. and Lubich, C.},
  booktitle={The Princeton Companion to Applied Mathematics},
  pages={293--305},
  year={2012},
  publisher={Princeton University Press}
}

@article{pmcmc,
  author = {C. Andrieu and A. Doucet and R. Holenstein},
  title = {Particle Markov chain Monte Carlo methods},
  journal = {Journal of the Royal Statistical Society: Series B (Statistical Methodology)},
  volume = {72},
  number = {3},
  pages = {269--342},
  year = {2010}
}

@article{andrieu2009pseudo,
  author = {C. Andrieu and G. O. Roberts},
  title = {The pseudo-marginal approach for efficient Monte Carlo computations},
  journal = {The Annals of Statistics},
  volume = {37},
  number = {2},
  pages = {697--725},
  month = {April},
  year = {2009}
}

@article{neal2012mcmc,
  title={MCMC using Hamiltonian dynamics},
  author={Neal, R. M.},
  journal={arXiv preprint arXiv:1206.1901},
  year={2012}
}

@article{creutz1988global,
  title={Global Monte Carlo algorithms for many-fermion systems},
  author={Creutz, M.},
  journal={Physical Review D},
  volume={38},
  pages={1228--1238},
  year={1988}
}

@article{nuts,
  title={The No-U-Turn sampler: adaptively setting path lengths in Hamiltonian Monte Carlo},
  author={Hoffman, M. D. and Gelman, A.},
  journal={Journal of Machine Learning Research},
  volume={15},
  number={1},
  pages={1593--1623},
  year={2014}
}

@article{hairer2003geometric,
  title={Geometric numerical integration illustrated by the St{\"o}rmer–Verlet method},
  author={Hairer, E. and Lubich, C. and Wanner, G.},
  journal={Acta Numerica},
  volume={12},
  pages={399--450},
  year={2003}
}

@article{titsias2018auxiliary,
  title={Auxiliary gradient-based sampling algorithms},
  author={Titsias, M. K. and Papaspiliopoulos, O.},
  journal={Journal of the Royal Statistical Society: Series B (Statistical Methodology)},
  volume={80},
  number={4},
  pages={749--767},
  year={2018}
}

@article{corenflos2024particle,
  title={Particle-MALA and Particle-mGRAD: Gradient-based MCMC methods for high-dimensional state-space models},
  author={Corenflos, A. and Finke, A.},
  journal={arXiv preprint arXiv:2401.14868},
  year={2024}
}

@article{nemeth2016particle,
  title={Particle metropolis-adjusted Langevin algorithms},
  author={Nemeth, C. and Sherlock, C. and Fearnhead, P.},
  journal={Biometrika},
  volume={103},
  number={3},
  pages={701--717},
  year={2016}
}

@article{poy,
  title={Particle approximations of the score and observed information matrix in state space models with application to parameter estimation},
  author={Poyiadjis, G. and Doucet, A. and Singh, S. S.},
  journal={Biometrika},
  volume={98},
  number={1},
  pages={65--80},
  year={2011}
}

@article{alenlov2021pseudo,
  title={Pseudo-marginal Hamiltonian Monte Carlo},
  author={Alenl{\"o}v, J. and Doucet, A. and Lindsten, F.},
  journal={Journal of Machine Learning Research},
  volume={22},
  number={141},
  pages={1--45},
  year={2021}
}

@inproceedings{chen2014stochastic,
  title={Stochastic gradient Hamiltonian Monte Carlo},
  author={Chen, T. and Fox, E. and Guestrin, C.},
  booktitle={International Conference on Machine Learning},
  pages={1683--1691},
  year={2014},
  publisher={PMLR}
}

@inproceedings{welling2011bayesian,
  title={Bayesian learning via stochastic gradient Langevin dynamics},
  author={Welling, M. and Teh, Y. W.},
  booktitle={Proceedings of the 28th international conference on machine learning (ICML-11)},
  pages={681--688},
  year={2011}
}

@book{liu2001monte,
  title={Monte Carlo Strategies in Scientific Computing},
  author={Liu, J. S.},
  year={2001},
  publisher={Springer},
  address={New York},
  note={MR1842342}
}

@techreport{neal1993probabilistic,
  title={Probabilistic inference using Markov chain Monte Carlo methods},
  author={Neal, R. M.},
  year={1993},
  institution={University of Toronto}
}

@article{duane1987hybrid,
  title={Hybrid monte carlo},
  author={Duane, S. and Kennedy, A. D. and Pendleton, B. J. and Roweth, D.},
  journal={Physics Letters B},
  volume={195},
  number={2},
  pages={216--222},
  year={1987},
  publisher={Elsevier}
}

@book{delmoral,
  author = {Del Moral, P.},
  title = {Feynman-Kac Formulae: Genealogical and Interacting Particle Systems with Applications},
  publisher = {Springer-Verlag},
  address = {New York},
  year = {2004}
}

@article{hmcexperiment3,
  author  = {Douc, R. and Garivier, A. and Moulines, E. and Olsson, J.},
  title   = {Sequential Monte Carlo Smoothing for General State Space Hidden Markov Models},
  journal = {The Annals of Applied Probability},
  volume  = {21},
  pages   = {2109--2145},
  year    = {2011}
}

@article{hmcexperiment2,
  author  = {Del Moral, P. and Doucet, A. and Singh, S. S.},
  title   = {A Backward Particle Interpretation of Feynman--Kac Formulae},
  journal = {ESAIM: Mathematical Modelling and Numerical Analysis},
  volume  = {44},
  pages   = {947--975},
  year    = {2010}
}

@inproceedings{Klaas2006,
  author    = {Klaas, M. and Briers, M. and de Freitas, N. and Doucet, A. and Maskell, S. and Lang, D.},
  booktitle = {Proceedings of the 23rd International Conference on Machine Learning - ICML '06},
  pages     = {481--488},
  year      = {2006},
  publisher = {ACM Press}
}

@techreport{hmcexperiment1,
  author      = {Del Moral, P. and Doucet, A. and Singh, S. S.},
  title       = {Forward Smoothing Using Sequential Monte Carlo},
  institution = {Cambridge University},
  number      = {638},
  type        = {Technical Report},
  year        = {2009},
  note        = {CUED-F-INFENG, Preprint}
}

@article{Osmundsen2018,
  author  = {Osmundsen, K. and Kleppe, T. S. and Liesenfeld, R.},
  title   = {Pseudo-Marginal Hamiltonian Monte Carlo with Efficient Importance Sampling},
  journal = {SSRN Electronic Journal},
  year    = {2018},
  note    = {Available at \url{https://ssrn.com/abstract=3304077}},
  doi     = {10.2139/ssrn.3304077}
}

@article{delmoral2,
  author = {F. C{\'e}rou and P. Del Moral and A. Guyader},
  title = {A non asymptotic variance theorem for unnormalized Feynman-Kac particle models},
  journal = {Annales de l'Institut Henri Poincar{\'e}},
  volume = {47},
  pages = {629--649},
  year = {2011}
}

@article{conc1,
  author = {N. Kantas and A. Doucet and S. Singh and J. Maciejowski and N. Chopin},
  title = {On Particle Methods for Parameter Estimation in State-Space Models},
  journal = {Statistical Science},
  volume = {30},
  pages = {328--351},
  year = {2015},
}

@article{nemeth2013particle,
  author       = {Nemeth, C. and Fearnhead, P. and Mihaylova, L.},
  title        = {Particle Approximations of the Score and Observed Information Matrix for Parameter Estimation in State Space Models with Linear Computational Cost},
  journal      = {arXiv preprint arXiv:1306.0735v1},
  year         = {2013},
}

@inproceedings{dahlin2013particle,
  author    = {Dahlin, J. and Lindsten, F. and Schön, T.B.},
  title     = {Particle Metropolis Hastings Using Langevin Dynamics},
  booktitle = {2013 IEEE International Conference on Acoustics, Speech and Signal Processing},
  pages     = {6308--6312},
  year      = {2013},
  publisher = {IEEE},
  month     = {May},
}

\end{document}